\documentclass{mn2e}
\input epsfig.sty
\newcommand{\be}{\begin{equation}}
\newcommand{\ee}{\end{equation}}
\newcommand{\Msun}{\mbox{$\rm M_{\odot}$}}
\newcommand\lsim{\mathrel{\rlap{\lower4pt\hbox{\hskip1pt$\sim$}}
    \raise1pt\hbox{$<$}}}
\newcommand\gsim{\mathrel{\rlap{\lower4pt\hbox{\hskip1pt$\sim$}}
    \raise1pt\hbox{$>$}}}
\newcommand\esim{\mathrel{\rlap{\raise2pt\hbox{\hskip0pt$\sim$}}
    \lower1pt\hbox{$-$}}}

\begin{document}
\journal{astro-ph/0305357}
\title[Cosmological perturbations and the reionization epoch]
{Cosmological perturbations and the reionization epoch}
\author[Pedro P. Avelino and Andrew R. Liddle]
{Pedro P. Avelino$^{1,2}$ and Andrew R. Liddle$^{2}$\\ 
$^1$Centro de F\'{\i}sica do Porto e Departamento de F\'{\i}sica da Faculdade 
de Ci\^encias da 
Universidade do Porto,\\ \hspace*{1cm} Rua do Campo Alegre 687, 4169-007, Porto, 
Portugal\\
$^2$Astronomy Centre, University of Sussex, Brighton BN1 9QJ, United Kingdom\\}
\maketitle
\begin{abstract}
We investigate the dependence of the epoch of reionization on the properties of 
cosmological perturbations, in the context of cosmologies permitted by WMAP.
We compute the redshift of reionization using a simple model based on the
Press--Schechter approximation. For a power-law initial spectrum we estimate 
that
reionization is likely to occur at a redshift $z_{{\rm reion}} = 17^{+10}_{-7}$,
consistent with the WMAP determination based on the
temperature--polarization cross power spectrum. We estimate the delay in 
reionization if there is a negative 
running of the 
spectral index, as weakly indicated by WMAP.
We then investigate the dependence of the
reionization redshift on the nature of the initial perturbations. We consider 
chi-squared probability distribution functions
with various degrees of freedom, motivated both by non-standard inflationary 
scenarios and by defect models. We find that in
these models reionization is likely occur much earlier, and to be a slower 
process, than in the case of 
initial gaussian fluctuations. We
also consider a hybrid model in which cosmic strings make an
important contribution to the seed fluctuations on scales relevant for
reionization.  We find that in order for that model to agree with the
latest WMAP results, the string contribution to the matter power
spectrum on the standard $8 h^{-1} {\rm Mpc}$ scale is likely to be at most
at the level of one percent, which imposes tight constraints on the
value of the string mass per unit length.

\end{abstract}
%%%%%%%%%%%%%%%%%%%%%%%%%%%%%%%%%%%%%%%%%%%%%%%%%%%%%%%%%%%%%%%%%%%%%%

\section{Introduction}

The reionization history of the Universe gives important insight into the 
epoch of structure formation in the Universe, with the initial reionization 
believed to have been caused by light from the first generation of massive stars 
(Couchman \& Rees 1986; Cen \& Ostriker 1992; Tegmark, Silk \& Blanchard 1994; 
Fukugita \& Kawasaki 1994). Over recent years, increasingly sophisticated 
techniques have been employed to estimate the reionization epoch in cosmological 
models (Haiman \& Loeb 
1997; Miralda-Escud\'e, Haehnelt \& Rees 2000; Cen 2002; Wyithe \& Loeb 2003; 
Somerville, Bullock \& Livio 2003; Fukugita \& Kawasaki 2003; see Barkana \& 
Loeb 2001 for a review).

Two important observations have begun to shed light on the reionization epoch, 
while painting a somewhat contradictory picture. The detection 
of
Gunn--Peterson troughs (Gunn \& Peterson 1965) in the absorption
spectra of distant quasars suggest a very late reionization for the Universe, at 
a redshift $z \sim 6$ (Becker et al.~2001; Fan et al.~2003; White et
al.~2003).  By contrast, the WMAP satellite's measurement of 
temperature--polarization correlations in the microwave background strongly 
suggest a higher optical depth from reionized electrons, implying 
that reionization took place at a higher redshift in the range from $z=11$ to 
$z=30$ (Kogut et al.~2003). It has been argued (Hui \& Haiman 2003) that such a 
high redshift of reionization would lead to an IGM temperature at $z \simeq 3$ 
which is lower than observations permit. In combination, these results suggest 
that the full history of ionization might be quite complicated (Hui \& 
Haiman 2003; Ciardi, Ferrara \& White 2003; Cen 2003) with more than one source 
of ionizing radiation.

Until recently, one of the key uncertainties in determining the reionization 
epoch was the choice of cosmological model, which determines the time of initial 
structure formation leading to the production of ionizing radiation. For 
example, Liddle \& Lyth (1995) carried out the first survey of cosmological 
parameter space and found a wide range of predictions for different models 
compatible with then-existing observations. These uncertainties have largely 
been removed by the establishment of a Standard Cosmological Model based on 
spatially-flat low-density Universe with inflationary perturbations. However, 
there remain significant uncertainties in the reionization epoch from the nature 
of the seed fluctuations. We will study two effects, one being a change in the 
short-scale power spectrum (due either to a running spectral index from 
inflation or from a topological defect contribution), and the other the effect 
of primordial non-gaussianity.

It is not our intention in this paper to provide a detailed modelling of the 
reionization history, but rather to introduce a highly-simplified model for 
reionization which allows us to assess how the epoch of reionization is modified 
if the cosmological model is altered. Our model is based on the Press--Schechter 
approach, following a strategy introduced by Tegmark et al.~(1994). 
Despite its simplicity, it makes predictions for the reionization epoch which 
are in broad agreement with those from deeper investigations using $N$-body/SPH 
simulations (Cen 2002, 2003; Ciardi et al.~2003) or more detailed 
semi-analytical modelling (Fukugita \& Kawasaki 1994, 2003; Wyithe \& Loeb 
2003). For the purpose of 
surveying cosmological models, we believe that our simple approach is well 
justified, and if accurate calculations become feasible for the standard 
cosmology in future, our results will enable a scaling to other cosmologies.

\section{The redshift of reionization}

In the context of Cold Dark Matter Models the first stars form in low
mass halos ($\sim 10^6 \Msun$) at high redshift. The precise
computation of the reionization redshift requires knowledge of the
fraction $f_{{\rm reion}}$ of baryons that must be bound in those halos
in order to bring the fraction of ionized gas close to $100 \%$. This
parameter is highly uncertain since it depends on several
astrophysical quantities, some of which are poorly known. Here we
shall use the estimates of $f_{{\rm reion}}$ derived by Tegmark et al.~(1994). 
They computed the fraction of the intergalactic
medium (IGM) that is ionized as a product of several factors, namely:
the fraction of baryons in non-linear structures, the number of UV
photons emitted into the IGM per proton in non-linear structures and
the net ionizations per emitted UV photon. They concluded that the
most reasonable estimate for the collapsed fraction required to induce
complete reionization is $f_{{\rm reion}} \sim 8 \times 10^{-3}$. This
estimate seems to be consistent with recent more detailed
investigations using large cosmological $N$-body/SPH simulations,
where radiative transfer calculations were carried out by embedding
massive Population III stars in gas clouds (Yoshida, Sokasian \&
Hernquist 2003). Tegmark et al.~(1994) also computed
conservative lower ($f_{{\rm reion}} \sim 4 \times 10^{-5}$)
and upper ($f_{{\rm reion}} \sim 0.8$) limits for $f_{{\rm
reion}}$ (in their terminology referred to as `optimistic' and `pessimistic') 
reflecting the large uncertainties associated with this
parameter. We shall use these estimates in the following sections in
order to estimate the redshift of reionization $z_{{\rm reion}}$ for
various models.

Despite the large uncertainties associated with the
determination of the collapsed baryon fraction necessary to reionize
the Universe, in the context of models with initial gaussian
fluctuations this fraction is exponentially sensitive to the
dispersion of the density field and hence to the redshift. This means
that a large uncertainty in $f_{{\rm reion}}$ translates into a
smaller uncertainty in the redshift of reionization, $z_{{\rm
reion}}$. However, we will see that this may not be the case when we
consider non-gaussian fluctuations.

\section{The mass fraction}

We shall use the Press--Schechter approximation (Press \& Schechter 1974) to 
compute the mass fraction, $f(>M)$ associated with collapsed objects
with mass larger than a given mass threshold $M$.  This was originally
proposed in the context of initial gaussian density perturbations and was much
later generalized to accommodate non-gaussian initial conditions (Chiu, Ostriker
\& Strauss 1998).  This generalization of the original Press--Schechter
approximation has successfully reproduced the results obtained from $N$-body
simulations with non-gaussian initial conditions (Robinson \& Baker 2000).
However, it has been shown by Avelino \& Viana (2000) that this generalization
does not adequately solve the cloud-in-cloud problem of Press--Schechter theory
and that deviations from the mass function predicted by this approach in the
rare events regime are expected (see also Inoue \& Nagashima 2002).  
Nevertheless,
given the large uncertainties in $f_{{\rm reion}}$, these deviations will have a
small impact on our final results and we shall not consider them further in the
following analysis.

In this context the mass fraction, $f(>M)$, is assumed to be
proportional to the fraction of space in which the linear density
contrast, smoothed on the scale $M$, exceeds a given threshold
$\delta_{{\rm c}}$: 
\be f(>M)= A_f \int_{\delta_{{\rm c}}}^\infty {\cal P}(\delta) d
\delta \,.  
\ee 
Here ${\cal P}(\delta)$ is the one-point probability
distribution function (PDF) of the linear density field $\delta$ and
$A_f$ is a constant which is computed by requiring that $f(>0)=1$, 
thus taking into account the accretion of material initially
present in underdense regions (note that $A_f=2$ in the case of gaussian
initial conditions).

Next we need to specify the filter function used to perform the smoothing, 
and the threshold value $\delta_{{\rm c}}$. We shall use a top-hat filter
\be
W(kR)=3\left(\frac{\sin (kR)}{(kR)^3}-\frac{\cos (kR)}{(kR)^2}\right)
\ee
where $M=4 \pi \rho_{{\rm b}} R^3/3$ (here $\rho_{{\rm b}}$ is 
the background matter 
density) and take $\delta_{{\rm c}}=1.7$, 
motivated by the spherical collapse model 
and $N$-body simulations. For spherical collapse $\delta_{{\rm c}}$ is 
almost independent of the background cosmology 
(e.g.~Eke, Cole \& Frenk 1996).

We consider an initial density field with a chi-squared one-point
probability distribution function (PDF) with $n$ degrees of freedom, 
the PDF having been shifted so that its mean is zero (such a PDF 
becomes gaussian when $n \to \infty$). We further
assume that the shape of the PDF is independent of the scale and
redshift under consideration, that is ${\cal P}_R(\delta) \equiv {\cal
P} (R,\delta/\sigma(R,z))/\sigma(R,z)$ is always the same function.
Although this is not expected to be precisely true in realistic
models, in particular those motivated by topological defects (Avelino
et al.~1998a), it is reasonable to expect it to be a good
approximation on the range of scales which are probed by reionization.

In order to compute the mass fraction $f(>M,z)$ associated with collapsed 
objects with mass larger than a given mass threshold $M$, as a function 
of redshift, one needs to compute the dispersion of the density field
\be
\sigma^2(R,0) = \int_0^\infty k^2 \left|{\delta _k}\right|^2 
W^2(kR)dk 
\ee
where $\left|{\delta _k}\right|^2$ is the power spectrum. 
For $z \gsim 1$ the dispersion of the density field is simply given by
\be
\sigma(R,z)=\frac{\sigma(R,0)}{g(\Omega_{{\rm m}}^0,\Omega_\Lambda^0)(1+z)},
\ee
where the suppression factor
\be
g(\Omega_{{\rm m}}^0,\Omega_\Lambda^0)=\frac{2.5 \Omega_{{\rm 
m}}^0}{(\Omega_{{\rm m}}^0)^{4/7}-\Omega_\Lambda^0+
(1+\Omega_{{\rm m}}^0/2)(1+\Omega_\Lambda^0/70)},
\ee
accounts for the dependence of the growth of density perturbations on the 
cosmological parameters $\Omega_{{\rm m}}^0$ and $\Omega_\Lambda^0$ (Carroll, 
Press 
\& Turner 1992). This has also been verified to be a good approximation in the 
context of generic models with topological defects 
(Avelino \& de Carvalho 1999).

\section{Results}

Throughout, we shall adopt a cosmological model motivated by the WMAP results 
(Bennett et 
al.~2003). As their mild evidence for a running of the spectral index is driven 
by lyman-$\alpha$ data, whose interpretation has proven controversial 
(e.g.~Seljak, McDonald \& Makarov 2003), we shall 
take as our base cosmology throughout their preferred model assuming a power-law 
initial spectrum (Spergel et al.~2003). The parameters are a matter density 
$\Omega_{{\rm m}}^0=0.29$, 
dark energy density $\Omega_\Lambda^0=0.71$,
baryon density $\Omega_{{\rm B}}^0=0.047$, Hubble parameter $h=0.72$, 
normalization $\sigma_8=0.9$, and perturbation spectral index $n_{{\rm 
s}}=0.99$.

To determine the amplitude of perturbations on the short scales relevant to 
reionization, for inflationary perturbations we use the transfer function from 
Bardeen et al.~(1986):
\be
\left|{\delta _k}\right|_I^2  \propto k^{n_s} \left[ \frac{\ln(1+\epsilon_0 
q)}{\epsilon_0 q} \left({ \sum_{i=0}^4(
\epsilon_i q)^i}\right)^{-1/4}\right]^2,
\ee 
where $q=k/h \Gamma$, $[k]={\rm{Mpc}}^{-1}$,
\be
\epsilon=[2.34, 3.89, 16.1, 5.46, 6.71],
\ee 
and
\be
\Gamma=\Omega_{{\rm m}}^0 h \exp \left[-\Omega_{{\rm 
B}}^0(1+\sqrt{2h}/\Omega_{{\rm m}}^0)\right] \,,
\ee
is the shape parameter (Sugiyama 1995). We note that 
this transfer function was shown to be a good approximation for wavelengths 
small enough to come into the horizon before matter--radiation equality if 
$\Omega_{{\rm B}}=0$ (Weinberg 2002). Also, although the effect of a non-zero 
baryon fraction is more complex than a simple rescaling of the 
shape parameter (Weinberg 2002), the Sugiyama correction turns out to 
be reasonably accurate for the values of the cosmological parameters we shall 
consider in this letter.

\subsection{The Standard Cosmological Model}

We begin by considering the simplest case, where the initial perturbations are 
gaussian. Computing the perturbations and mass fraction as described, following 
Tegmark et al.~(1994) and Liddle \& Lyth (1995), we find a central value of 
$z_{{\rm reion}} = 17$, with the range from highest to lowest plausible values 
being from 27 down to 10. This range is in excellent agreement with the WMAP 
determination of the reionization epoch from the temperature--polarization 
correlation function. Based on the assumption of instantaneous and complete 
reionization, Kogut et al.~(2003) find $z_{{\rm reion}} = 17 \pm 3$ (one-sigma) 
based on the cross-correlation alone, while a global fit to WMAP and other data 
(Spergel et al.~2003) gives $z_{{\rm reion}} = 17 \pm 4$. However this model 
appears in conflict with the high-redshift quasar data, and Kogut et al.~(2003) 
show that allowing more complex ionization histories can widen the allowed range 
considerably (see also Haiman \& Hui 2003; Ciardi et al.~2003; Cen 2003).

The predictions above are our fiducial result, against which we will be 
comparing variations under different assumptions about the initial 
perturbations. If more detailed calculations of the reionization epoch give a 
different value, our estimates of the fractional variation about this value in 
other 
cosmologies should remain valid to a good approximation. For instance, a recent 
detailed analysis by Fukugita \& Kawasaki (2003) suggests a somewhat lower value 
of $z_{{\rm reion}} = 13$ as the earliest possible reionization, with 
the reionization optical depth receiving a significant contribution before 
reionization is complete.

\subsection{Running spectral index}

While our main focus is to examine the effects of primordial 
non-gaussianities, 
we begin with a brief exploration of the effect of a possible running of the 
spectral index, motivated by WMAP. The WMAP preferred value $dn_s/d\ln k = 
-0.031$ (Spergel et al.~2003)
is much larger than anticipated from simple inflation models (Kosowsky \& Turner 
1995; Copeland, Grivell \& Liddle 1998), and is 
large enough to have significant effect. The WMAP running model has been studied 
in detail by Somerville et al.~(2003); we will consider a general running.

\begin{figure}
\centering
\psfig{figure=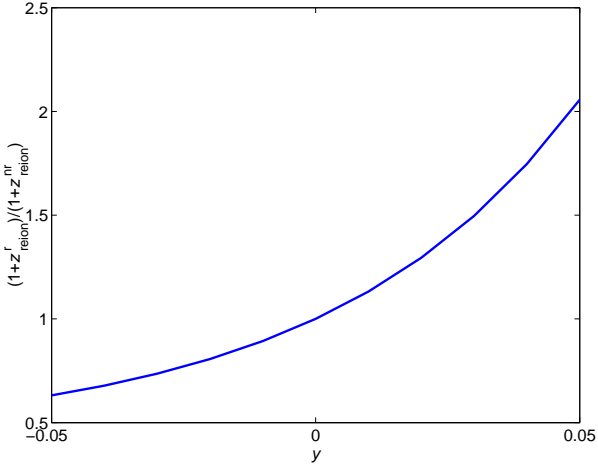,width=8cm}
\caption[Figure1]{\label{f:run} The shift in the expected epoch of reionization 
induced by the inclusion of running. Negative running, as mildly indicated by 
WMAP, reduces the redshift of reionization.}
\end{figure}

To determine the effect of running, we define it at the scale $k=0.0072 \, {{\rm 
Mpc}}^{-1}$ at which the WMAP running cosmology has the same spectral index as 
the power-law fit above. A negative running then progressively reduces 
short-scale power, delaying reionization, while a positive running has the 
opposite effect. The shift in the reionization epoch is shown in 
Figure~\ref{f:run}, and is well fit (to around five per cent within the region 
plotted) by
\be
\frac{1+z^{\rm r}_{\rm reion}}{1+z^{\rm nr}_{\rm reion}} = 1+14y+138  y^2
\quad ; \quad y \equiv dn_s/d\ln k \,,
\ee
where `nr' indicates the fiducial model without running, and `r' the inclusion 
of running.
We conclude that running can systematically shift the expected redshift of 
reionization. For $dn_s/d\ln k = -0.031$ the shift in 
$1+z_{\rm reion}$ is about $30 \%$, which is consistent with the shift found by 
Somerville et al.~(2003) for the WMAP running model. Since running at that level 
is permitted by the current data, we conclude that uncertainty in the 
small-scale power spectrum contributes a 
fractional uncertainty in predicting $z_{{\rm reion}}$ of at least that level, 
unless one imposes a prior restricting the effect of running.

\subsection{Non-gaussian inflation}

We now consider two possible models of non-gaussian primordial
perturbations.  The first is a model with primordial fluctuations which may have
been generated during inflation, but featuring a chi-squared probability
distribution as may appear in the context of some non-standard inflationary
models (see for example Salopek 1992; Linde \& Mukhanov 1997; Peebles 1999; 
Bartolo, Matarrese \& Riotto 2001, 2002; Bernardeau \& Uzan 2003).  

\begin{figure}
\centering
\psfig{figure=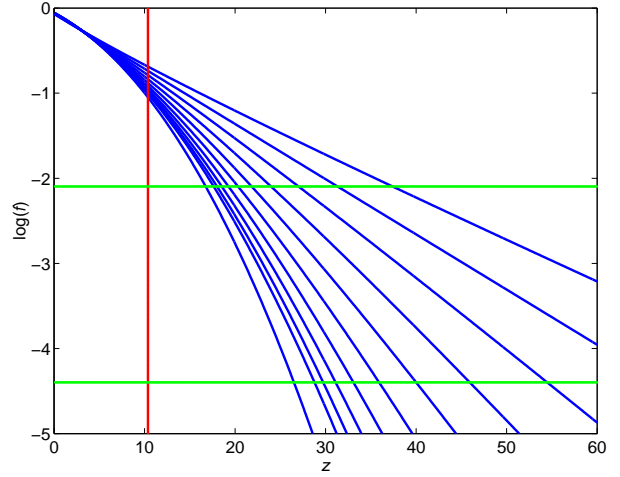,width=8cm}
\caption[Figure1]{\label{f:inf} The evolution of the mass fraction 
$f(>10^6 \, \Msun)$ as a function of redshift $z$ for non-gaussian inflation,
as a function of the number of chi-squared degrees of freedom 
($n=1$, $2$, $4$, $8$, $16$, $32$, $64$, $128$, $256$, $512$, $\infty$ from top 
to bottom). 
The intersection of mass fraction curves with the horizontal lines give  
the upper limit and best estimate for the redshift of 
reionization. The lower limit is given by the vertical line.}
\end{figure}

In Figure~\ref{f:inf} we plot the evolution of the mass fraction $f(>10^6
\Msun)$, as a function of redshift $z$, for chi-squared distributions with 
different numbers of degrees of freedom 
($n=1,2,4,8,16,32,64,128,256,512,\infty$). The intersections of the mass
fraction curves with the horizontal lines give the upper limit and
best estimate for the redshift of reionization. 

The lower limit
is instead given by the vertical line. The reason is that in this case $f_{\rm 
reion} \sim 0.8$ is too large to be reliably predicted by Press--Schechter, so  
we follow Liddle \& Lyth (1995) in replacing this 
condition by the 
criterion that a significant fraction of the mass has to have collapsed 
for reionization to occur. We quantify this assuming a crude lower
estimate for the redshift of reionization given by 
$\sigma(10^6 \, \Msun , z_{\rm reion})=1$ that leads to
$$
1+z_{\rm reion}=\sigma(10^6 \, \Msun,0)/g(\Omega_{{\rm m}}^0, \Omega_\Lambda^0),
$$
which is independent of the degree of non-gaussianity of the initial 
density field.
 
We see that (except in the extreme case of the lower limit) the predicted
redshift of reionization increases as the number of degrees of freedom of the
chi-squared distributed PDF decreases, as does the uncertainty on the redshift 
of
reionization.  We also see that the evolution of $f(> 10^6 \, \Msun)$ with
redshift is not as fast for smaller $n$, which indicates that reionization may 
be
a slower process.  This happens because in this case $f(> 10^6
\, \Msun)$ is less sensitive to the dispersion of the density field than in the
gaussian case.

We cannot exclude any of these models, as the lower estimate for the redshift is 
independent of the degree of non-gaussianity. However, the best estimate 
increases as the number of degrees of freedom reduces, becoming 
inconsistent with WMAP at low $n$. If in the future $f_{{\rm reion}}$ can be 
accurately 
determined, our results will quickly indicate where the limit sets in.

\subsection{Inflation plus defects}

Our second non-gaussian model is one in which both
string and inflationary perturbations are present. 
Although standard topological defect models are completely excluded as the sole
source of perturbations in the Universe, it remains plausible that
they may play a sub-dominant role (see for example Durrer, Kunz \& Melchiorri 
(2002) and refs therein). In particular, the popular hybrid
inflation scenario relies on a phase transition to end inflation,
which would be predicted to produce defects at a sub-dominant
level (Copeland et al.~1994; Contaldi, Hindmarsh \& Magueijo 1999). As we shall 
see, reionization is a particularly powerful probe
of the effects of defects. We will focus on inflation plus cosmic
strings, assuming initially that the two perturbation types have equal amplitude 
at
the standard $8 h^{-1} {\rm Mpc}$ scale.

A good fit to the string-seeded CDM power spectrum is given by 
(Wu et al.~2002)
\be
\left|{\delta _k}\right|_S^2  \propto (0.7q)^{p(q)},
\ee
where
\be
p(q)=0.9-\frac{2.7}{1+(2.8q)^{-0.44}}.
\ee
Given that the inflationary and string-induced perturbations are uncorrelated, 
we can write the combined power spectrum as
\be
\left|{\delta _k}\right|_{{\rm combined}}^2  \propto \alpha 
\left|{\delta _k}\right|_{I}^2 + (1-\alpha) \left|{\delta_k}\right|_S^2 \,,
\ee
where $\alpha$ is such that inflationary and string induced perturbations 
make an equal contribution to the dispersion of the density field, 
$\sigma_8$ on the standard $8 h^{-1} {\rm Mpc}$ scale. On the much
smaller scales relevant for reionization, the cosmic string
induced perturbations completely dominate over the inflationary ones.

The cosmic string seeded density perturbations on a given scale $R$ 
can be roughly divided into a nearly gaussian component plus a strongly-skewed 
non-gaussian part generated when the string correlation 
length was smaller/larger than $R$ respectively. Using simulation results for 
string-induced CDM perturbations, Avelino, Shellard \& Wu (2000) 
found that the positive side of the one-point PDF was well approximated 
by a chi-squared distribution with the number of degrees of freedom being 
an increasing function of scale. Although these simulations did not have 
enough dynamical range to probe the scales relevant to reionization, 
they showed clearly that on length scales smaller than\ $1.5 
{(\Omega_m^0 h^2)}^{-1}{\rm Mpc}$  the 
perturbations seeded by cos\-mic strings have a strong non-gaussian character 
(Avelino 
et al.~1998a).

\begin{figure}
\centering
\psfig{figure=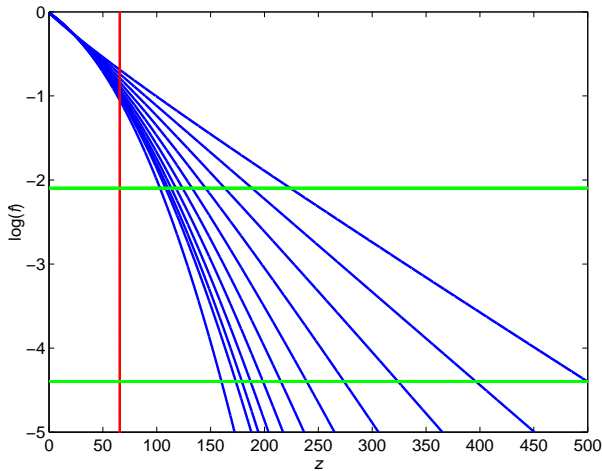,width=8cm}
\caption[Figure1]{\label{f:def} As Figure~\ref{f:inf}, 
but for a hybrid model with both string and 
inflationary perturbations. 
The two perturbation types are assumed 
to have equal amplitude at the standard $8 h^{-1} {\rm Mpc}$ 
scale. We treat the degree of non-gaussianity of the combined perturbation 
field as a variable parametrized by $n=1,2,4,8,16,32,64,128,256,512,\infty$ 
(the number of degrees of freedom of the assumed chi-squared one-point PDF).
Again, the intersection of mass fraction curves with the horizontal lines 
give the upper limit and best estimate for the redshift of 
reionization. The lower limit is given by the vertical line.}
\end{figure}

Our results are shown in Figure~\ref{f:def}. Here, we separate the effect of 
the non-gaussian nature of the string 
perturbations from that of the power spectrum. The power spectrum of 
string-induced CDM perturbations dominates on small scales due to the 
wake-like signature of cosmic strings leading to a small-scale power spectrum 
close to $k^{-2}$. Again the redshift of reionization 
increases as the number of degrees of freedom of the chi-squared distributed 
PDF decreases, the same effect as that obtained for non-gaussian inflation. 
As previously discussed we expect the results obtained 
for small $n$ to be representative of what is expected for the defect 
contribution (in particular cosmic strings).

In this case we see that the much larger amplitude of the 
cosmic string contribution to the matter power spectrum (relative to 
the inflationary one) on scales of the order of $10^6 \Msun$ 
results in a much earlier reionization ($z \gsim 65$). Even if the perturbations 
were gaussian (the lowest curve) we have very early reionization due to the 
different power spectrum shape, with non-gaussianity further exacerbating this. 
The predicted reionization epoch is completely 
incompatible with the observed microwave anisotropies, easily excluding models 
where the defects produce half the power at $8 h^{-1} \, {\rm Mpc}$. 
We note that in this case the string perturbation spectrum has a 
$\sigma^S_8=\sigma_8/\sqrt{2} \sim 0.64$ which is close to the expected 
value for a string mass per unit length $G \mu \sim 10^{-6})$ 
(Avelino et al.~1998b; Wu et al.~2002). Assuming the best guess for 
$f_{{\rm reion}}$,  
reionization seems to occur about a factor of $10$ earlier in redshift in the 
case of the hybrid model considered above (the exact number depending on 
the degree of non-gaussianity on the scales relevant for reionization). 
Accordingly, 
in order to restore viability with the data the string contribution has to be 
significantly lowered, at least by a factor of $10$. This  
indicates that the string contribution to the matter power 
spectrum on the standard $8 h^{-1} {\rm Mpc}$ would have to be at most at 
the level of one percent, in which case an upper limit to the string mass 
per unit length would be $G \mu \lsim 10^{-7}$. However we note that this 
result is sensitive 
to the highly-uncertain value of $f_{{\rm reion}}$. Still, we can 
clearly see that reionization imposes very strong constraints 
on any contribution from defects to the seed fluctuation spectrum.  We expect 
our overall results to remain valid at some level in the context of other defect 
models, as the small-scale non-gaussianity and excess power are generic 
predictions of models of this type (Pen, Spergel \& Turok 1994).

\section{Conclusions}

We have shown that in a model with initial gaussian fluctuations
reionization is expected to occur at a redshift consistent with the
recent WMAP results, while running of the spectral index within the range 
permitted by observations leads to a significant uncertainty in predicting the 
reionization redshift.
Although this result does not leave much room for
non-gaussian perturbations, we have shown that even a small level of
non-gaussianity may have interesting consequences. In the case of
an initial density field with a chi-squared PDF, motivated both by some
non-standard models of inflation and topological defects, reionization
is likely to be a slower process.  We also show that the 
reionization history of the Universe
imposes strong constraints on the energy scale of defects given their
ability to influence it through the induced small-scale excess power
and non-gaussianity. We estimate that in the case of a hybrid model
with cosmic strings the string contribution to the total power
spectrum on the standard $8 h^{-1} {\rm Mpc}$ is likely to be at most
at the level of one percent thus requiring a low value of the string
mass per unit length ($G \mu \lsim 10^{-7}$).

\section*{ACKNOWLEDGMENTS}
P.P.A. was partially supported by Funda{\c c}\~ao para a Ci\^encia e a
Tecnologia (Portugal) under contract CERN/POCTI/49507/2002. A.R.L.~was supported 
in part by the Leverhulme Trust.

\bsp

\end{document}